\documentclass[pra,twocolumn,showpacs,preprintnumbers,amsmath,amssymb,aps]{revtex4}
\usepackage{color}
\usepackage{bm, graphicx, amsmath}

\begin{document}
\title{A group-theoretic approach to elimination measurements of qubit sequences}
\author{Mark Hillery$^{1,2}$, Erika Andersson$^{3}$, and Ittoop Vergheese$^{3}$}
\affiliation{$^{1}$Department of Physics and Astronomy, Hunter College of the City University of New York, 695 Park Avenue, New York, NY 10065 USA \\ 
$^{2}$Graduate Center of the City University of New York, 365 Fifth Avenue, New York, NY 10016 \\
$^{3}$SUPA, Institute for Photonics and Quantum Sciences, School of Engineering and Physical Sciences, Heriot-Watt University, Edinburgh EH14 4AS, United Kingdom}

\begin{abstract}
Most measurements are designed to tell you which of several alternatives have occurred, but it is also possible to make measurements that eliminate possibilities and tell you an alternative that did not occur.  Measurements of this type have proven useful in quantum foundations and in quantum cryptography.  Here we show how group theory can be used to design such measurements.  After some general considerations, we focus on the case of measurements on two-qubit states that eliminate one state.  We then move on to construct measurements that eliminate two  three-qubit states and four four-qubit states.  A condition that constrains the construction of elimination measurements is then presented.  Finally, in an appendix, we briefly consider the case of elimination measurements with failure probabilities and an elimination measurement on $n$-qubit states.
\end{abstract}

\maketitle

\section{Introduction}
Measurements are usually made to identify one out of a number of possibilities.  In particular, in state discrimination measurements, in which the object is to determine which quantum state one has, the task is to design a measurement that will tell you exactly that, what the state of the system you have been given is.  Another possibility, however, is to design a measurement that tells you which state you do not have, i.e.\ one that eliminates a possibility.  A simple example of this is 
a measurement that eliminates one of the trine 
states of a qubit.  The trine states are $|0\rangle$, $(1/2) (-|0\rangle + \sqrt{3}|1\rangle )$, and $(-1/2) (|0\rangle + \sqrt{3}|1\rangle )$.  
The ``anti-trine states", which are orthogonal to the trine states, are $|1\rangle$, $(-1/2) (\sqrt{3}|0\rangle + |1\rangle )$, and $(1/2)(\sqrt{3} |0\rangle - |1\rangle)$. Suppose you are given a qubit, which is guaranteed to be in one of the trine states.  It is not possible to find a measurement that will definitely tell you which state you have, some probability of error or failure is necessary. But by making use of a POVM whose elements are proportional to projections onto the anti-trine states you can definitely find out a state you do not have.  For example, if you find a result corresponding to the state $|1\rangle$, you have not been given the state $|0\rangle$.

There has not been a great deal of work on state elimination measurements.  None the less, they have found application in studies of the hidden subgroup problem \cite{hoyer}, quantum foundations \cite{PBR,Caves2002}, quantum communication \cite{Perry}, and quantum cryptography \cite{Collins,RyanOT}.  Perhaps the most extensive study so far is by Bandyopadhyay \emph{et al.}, who applied semi-definite programming to an examination of single-state elimination measurements \cite{Bando2014}.  Single state elimination also goes under the name ``anti-distinguishability'', and the focus of the work in that area has been in finding conditions for determining when a set of states is anti-distiguishable, i.e. when does there exist a POVM each of whose outcomes corresponds to eliminating one of the states \cite{Caves,Heinosaari1,Havlicek}.  Recently, a connection between anti-distinguishability and non-contextuality inequalities was found \cite{Leifer}.  Measurements for eliminating pairs of two qubit states, which generalize some of the results in \cite{PBR}, were presented in \cite{Crickmore}.

Here we are interested in looking at measurements that eliminate sets of states, not just single states as in anti-distinguishability.  We will focus on states that are qubit sequences, where each qubit can be in one of several different states..  Previous studies of elimination measurements for states of this type considered only two states per qubit, and the elimination of a single state of the sequence.  We will construct measurements that remove both of these restrictions.  The tool we will use is group theory, and we will use it to construct covariant elimination measurements \cite{Dariano}.  Group theory was used to generate sets of anti-distinguishable states in the case that the representation of the group is irreducible in \cite{Heinosaari1}.  We will go beyond this and consider the case in which the representations is reducible, but with the restriction that each irreducible representation appears at most once.  This turns out to be quite adequate  for producing elimination measurements of qubit sequences.  We will start with measurements that eliminate one state, but then go on to find measurements that eliminate two or four states.  Finally, we find a condition that places a restriction on what types of elimination measurements are possible.

\section{Group theory preliminaries}
Suppose one has a collection of states, $\{ |\psi_{g}\rangle = \Gamma (g) |\psi_{e}\rangle\, | \, g\in G \}$.  Here, $G$ is a group, $\Gamma (g)$ for $g\in G$ is a unitary representation of the group in which each irreducible representation appears at most once, and $e\in G$ is the identity element of the group.  We want to find a POVM, $\{ \Pi_{g}\, | \, g\in G\}$, for which each element corresponds to eliminating one of the states in the set.  That means that, if $\Pi_{g}$ is the element corresponding to $g$, then $\Pi_{g}|\psi_{g}\rangle = 0$.  Let us assume for now that the POVM elements are rank one, and can be expressed in the form
\begin{equation}
\label{povm} 
\Pi_{g}= \Gamma (g) |X\rangle\langle X| \Gamma (g)^{-1} ,
\end{equation}
where $|X\rangle$ is a vector yet to be determined.  Note that if $\Pi_{e} |\psi_{e}\rangle =0$, which is equivalent to the condition $\langle X|\psi_{e}\rangle = 0$, then we will have $\Pi_{g}|\psi_{g}\rangle = 0$.

The representation $\Gamma$ can be decomposed into irreducible representations, $\Gamma_{p}$,
\begin{equation}
\Gamma = \bigoplus_{p} \Gamma_{p} ,
\end{equation}
where each $\Gamma_{p}$ acts on an invariant subspace, that is, each $\Gamma_{p}(g)$, for $g\in G$, maps the subspace into itself.  Let $P_{p}$ be the projection onto the subspace corresponding to $\Gamma_{p}$. Then a theorem from group representation theory implies that
\begin{equation}
\label{decomp}
\frac{1}{|G|} \sum_{g\in G} \Gamma (g) |X\rangle\langle X| \Gamma (g)^{-1} = \sum_{p} \frac{1}{d_{p}} \| X_{p}\|^{2} P_{p} ,
\end{equation}
where $d_{p}$ is the dimension of the representation $\Gamma_{p}$, $|G|$ is the number of elements in $G$, and $|X_{p}\rangle = P_{p}|X\rangle$.  Now suppose we find a vector $|X\rangle$ in Eq.\ (\ref{povm}) that satisfies $\| X_{p}\|^{2} = d_{p}/|G|$.  We will then have that $\sum_{g} \Pi_{g} = I$, as is required for the elements of a POVM.

As a simple example, let us apply this to a single qubit with the group $\mathbb{Z}_{3} =\{ e,g,g^{2}\}$, where $e$ is the identity element and $g^{3}=e$.  We shall choose a two-dimensional representation of $\mathbb{Z}_{3}$, and identify $g$ with the matrix (in the computational basis)
\begin{equation}
V=\left( \begin{array}{cc} -1/2 & \sqrt{3}/2 \\ -\sqrt{3}/2 & -1/2 \end{array} \right) ,
\end{equation}
that is, $\Gamma (g) = V$, which is a rotation by $2\pi /3$ in the $x-y$ plane.  This matrix has eigenstates
\begin{equation}
|u_{\pm}\rangle = \frac{1}{\sqrt{2}}\left( \begin{array}{c} 1 \\ \pm i \end{array} \right)  ,
\end{equation}
corresponding to eigenvalues $e^{\pm 2\pi i/3}$, respectively.  We will be finding an elimination measurement for the vectors $V^{j}|0\rangle$, where $j \in \{ 0,1,2 \}$.  Note that the vectors $|0\rangle$, $V|0\rangle$, and $V^{2}|0\rangle$ are just the trine states.

The vectors corresponding to the invariant subspaces of the irreducible representations are $|u_{j}\rangle$, where $j = \pm$.  The vector $|X\rangle$ is now
\begin{equation}
|X\rangle = \frac{1}{\sqrt{3}} \sum_{j=\pm} e^{i\phi_{j}}|u_{j}\rangle ,
\end{equation}
which follows from the fact that $\| X_{p}\|^{2}=d_{p}/|G|$, $d_{p}=1$, and $|G|=3$.
The condition that $\langle 0|X\rangle = 0$ is
\begin{equation}
\sum_{j=\pm} e^{i\phi_{j}} = 0.
\end{equation}
This equation is easy to satisfy with the obvious choice of  $\phi_{+} = 0$ and $\phi_{-} = \pi$.  Making this choice we find that $|X\rangle =i\sqrt{2/3} |1\rangle$.  The POVM elements are found by applying $V^{j}$, where $j \in \{ 0,1,2 \}$ to $|X\rangle$, and they are proportional to projections onto the anti-trine states.

\section{Single-state elimination for two qubits}

Let's now apply the group theory perspective to the measurement of four two-qubit states, which will be a generalisation of the measurement considered in \cite{PBR}.  We will make use of the group $\mathbb{Z}_{2}$.  This group has two elements, $\mathbb{Z}_{2}=\{ e,g\}$, where $e$ is the identity element and $g^{2}=e$.  We will choose a two-dimensional representation of this group acting in the qubit space spanned by the computational basis vectors $|0\rangle$ and $|1\rangle$.  The representation is specified by $\Gamma (e)=I$, the identity operator, and $\Gamma (g) = R$, the reflection through the $x$ axis,  $R|0\rangle = |0\rangle$ and $R|1\rangle = - |1\rangle$.  The two irreducible representations of  $\mathbb{Z}_{2}$ are $\Gamma_{1}(e) =1$, $\Gamma_{1}(g)=1$, and $\Gamma_{2}(e) =1$, $\Gamma_{2}(g)=-1$.  The invariant subspace corresponding to $\Gamma_{1}$ is spanned by $|0\rangle$, and the invariant subspace corresponding to $\Gamma_{2}$ is spanned by $|1\rangle$.  For the state $|\psi_{e}\rangle$, we shall choose
\begin{equation}
|+ \theta\rangle = \cos\theta |0\rangle + \sin\theta |1\rangle ,
\end{equation}
where $0 \leq \theta \leq \pi /4$.  This state is mapped into the state 
\begin{equation}
|- \theta\rangle = \cos\theta |0\rangle - \sin\theta |1\rangle
\end{equation}
by $R$.

Going now to two qubits, our task is to find a measurement to eliminate one of the four states $|\pm \theta\rangle \otimes|\pm \theta\rangle$, where the pluses and minuses for each qubit are independent.  Pusey et al. gave such a measurement for $\theta=\pi/8$. Our group is now $\mathbb{Z}_{2}\times \mathbb{Z}_{2}$ and the representation, $\Gamma (g)$, is now $\{ I\otimes I, I\otimes R, R\otimes I, R\otimes R \}$, which is a four-dimensional representation.  The irreducible representations are just the products of the irreducible representations for $\mathbb{Z}_{2}$, and there are four of them.  The invariant subspaces corresponding to the irreducible representations of $\mathbb{Z}_{2} \times \mathbb{Z}_{2}$ are just $|j\rangle \otimes|k\rangle$, where $j,k \in \{ 0,1 \}$.  The vector $|\psi_{e}\rangle$ is now $|+\theta\rangle \otimes  |+\theta\rangle$.  The vector $|X\rangle$ can be chosen to be
\begin{equation}
\label{XPBR}
|X\rangle = \frac{1}{2} \sum_{j,k=0,1} e^{i\phi_{jk}}|j\rangle \otimes|k\rangle  .
\end{equation}
We can set $\phi_{00} = 0$ without loss of generality.  The condition that $\langle +\theta, +\theta |X\rangle = 0$ is then
\begin{equation}
\cos^{2}\theta + e^{i\phi_{11}} \sin^{2}\theta + (e^{i\phi_{01}} + e^{i\phi_{10}})  \cos\theta \sin\theta =0 .
\end{equation}
Dividing through by $\cos^{2}\theta$ we get
\begin{equation}
1 + e^{i\phi_{11}} \tan^{2}\theta + (e^{i\phi_{01}} + e^{i\phi_{10}})  \tan\theta = 0 .
\end{equation}
Let's make the Ansatz $\phi_{11}=\pi$ and $\phi_{01}=\phi_{10}=\phi + \pi$.  This gives us
\begin{equation}
1- \tan^{2}\theta -2\tan\theta \cos\phi = 0 .
\end{equation}
For this to have a solution, it must be the case that
\begin{equation}
1- \tan^{2}\theta -2\tan\theta \leq 0 ,
\end{equation}
and this will be true if $\tan\theta \geq \sqrt{2}-1$, i.e.\ $\theta \geq \pi /8$.  For $\theta$ satisfying this condition, we have the POVM whose elements are $\Gamma(g)|X\rangle\langle X|\Gamma^{\dagger}(g)$ for $g \in \mathbb{Z}_{2} \times \mathbb{Z}_{2}$.  Each element corresponds to eliminating one of the four states.  Note that while the states we are considering are separable, the POVM elements are projections onto entangled states.  All of this is consistent with the results in \cite{PBR}, where the elimination measurement was given for $\theta = \pi/8$.  Since we started with the assumption that the POVM elements are rank 1, we have strictly speaking not proven that a measurement that always eliminates one two-qubit state is impossible for $\theta<\pi/8$, but it turns out that this is the case \cite{Crickmore}.  As we will see in Appendix B, for $\theta <\pi/8$ it is possible to sometimes eliminate one two-qubit state, that is, the measurement may sometimes fail, but when it succeeds, it will conclusively eliminate one state..
Crickmore et al. \cite{Crickmore} give an alternative construction for the measurement in the whole range $0\ <\theta\le \pi/4$, and also prove optimality.

Now suppose we want to consider more states.  In particular, we would like to consider $N$ states for each qubit.  First we need the group $\mathbb{Z}_{N} = \{ g^{j}\, | \,  j=0,1,2,\ldots N-1 \}$, where $g^{0}=g^{N}=e$, and its irreducible representations, which are given by $\Gamma_{k}(e) = 1$ and $\Gamma_{k}(g) = \exp (2k\pi i/N)$ for $k=0,1,2,\ldots N-1$.  We will choose the representation $\Gamma (e)=I$ and $\Gamma (g)= S_{N}$, where $S_{N}|0\rangle = |0\rangle$ and $S_{N}|1\rangle = \exp (2\pi i /N) |1\rangle$.  This representation is a direct sum of $\Gamma_{0}$ and $\Gamma_{1}$, and the invariant subspaces, as before, are spanned by $|0\rangle$ and by $|1\rangle$.  For two qubits, the group is $\mathbb{Z}_{N} \times \mathbb{Z}_{N}$.  The vectors to be measured are generated by applying $S_{N}^{j}\otimes S_{N}^{k}$, for $j,k \in \{ 0,1,2,\ldots N-1 \}$ to $|+\theta\rangle \otimes|+\theta\rangle$, and the POVM elements are generated by applying these same operators to the vector $|X\rangle$, which is the same, up to a factor, as above.  In particular, the factor of $1/2$ in Eq.\ (\ref{XPBR}) will be replaced by $1/N$, since the group now has $N^{2}$ rather than $4$ elements.  Therefore, we can see that the group theory gives us an elimination measurement for a larger set of states with almost no additional work.  Note that the POVM elements are proportional to projections onto entangled states.

\section{A non-abelian group}
Up until now we have only made use of abelian groups,  so now let us look at a non-abelian group.  A simple non-abelian group is the dihedral group $D_{3}$, which consists of rotations and reflections in the plane that leave an equilateral triangle invariant.  It has six elements, $\{ e,r,r^{2}, s, rs, r^{2}s \}$, where $r^{3}=e$ and $s^{2}=e$.  The dihedral group $D_3$ is isomorphic to the symmetric group~$S_3$, i.e., the group of permutations of three elements. The mapping is defined  by $s\mapsto(12)$, $r\mapsto(123)$.
The group has three conjugacy classes $C_{e}=\{ e\}$, $C_{r}=\{ r,r^{2}\}$, and~$C_{s}=\{ s, rs, r^{2}s \}$.  It has three irreducible representations, $\Gamma_p$ for $p=1,2,3$, where $\Gamma_1$ and $\Gamma_2$ are one-dimensional and $\Gamma_3$ is two dimensional.  The character table for the group is given in Table~\ref{t-1}.

\begin{table}
\centering
\begin{tabular}{|c|c|c|c|} \hline
 & $C_{e}$ & $C_{r}$ & $C_{s}$ \\ \hline $\Gamma_1$ & $1$ & $1$& $1$ \\ \hline $\Gamma_2$ & $1$ & $1$ & $-1$ \\ \hline $\Gamma_3$ & $2$ & $-1$ & $0$ \\ \hline 
\end{tabular}
\caption{\label{t-1}Character table for $D_{3}$.}
\end{table}

The one-dimensional representations are the trivial representation, $\Gamma_1(g)=1$ for all $g\in D_3$, and the so-called sign or alternate representation, defined by $
\Gamma_2(r)=1$ and $\Gamma_2(s)=-1$ for the generators of the group $r$ and $s$. 
For the representation $\Gamma_3$, we can take the matrices
\begin{equation}
\Gamma_3(r)=\left( \begin{array}{cc} -1/2 & -\sqrt{3}/2 \\ \sqrt{3}/2 & -1/2 \end{array} \right),\ \Gamma_3(s)=\left( \begin{array}{cc} 1 & 0 \\ 0 & -1 \end{array}\right) ,
\end{equation}
 expressed in the computational basis $\{ |0\rangle ,|1\rangle \}$. 

Suppose we have two qubits, which transform according to the representation $\Gamma_3\otimes \Gamma_3$, that is $\Gamma (g) = \Gamma_{3}(g) \otimes \Gamma_{3}(g)$ for $g\in D_{3}$.  For $|\psi_{e}\rangle$ we will choose $|0\rangle \otimes |+x\rangle$, where $|\pm x\rangle = (|0\rangle \pm |1\rangle )/\sqrt{2}$, and the application of $\Gamma (g)$ to this state for the different possible values of $g$ yields a set of $6$ product states in a four dimensional space.  We now want to find a POVM that eliminates one of these states, which means we want to find a suitable vector $|X\rangle$.   

The product representation, $\Gamma$ can be decomposed into irreducible representations, 
\begin{equation}
\Gamma_3\otimes \Gamma_3=\Gamma_1 \oplus \Gamma_2 \oplus \Gamma_3 .
\label{D3rep}
\end{equation}
For the invariant subspaces, we find that  $|v_{1}\rangle = (|00\rangle + |11\rangle )/\sqrt{2}$ transforms as $\Gamma_{1}$, $|v_{2}\rangle = (|01\rangle - |10\rangle )/\sqrt{2}$ transforms as $\Gamma_{2}$, and the subspace that transforms as $\Gamma_{3}$ is spanned by $|v_{3}\rangle = (|00\rangle - |11\rangle )/\sqrt{2}$ and $|v_{4}\rangle = (|01\rangle + |10\rangle )/\sqrt{2}$.  In terms of these states, we have
\begin{equation}
|0\rangle \otimes |+x\rangle = \frac{1}{2} \sum_{j=1}^{4} |v_{j}\rangle .
\end{equation}
The vector $|X\rangle$ must be orthogonal to this vector and satisfy $\|X_{1}\|^{2}=\|X_{2}\|^{2}=1/6$ and $\|X_{3}\|^{2}=1/3$.  We find that
\begin{equation}
|X\rangle = \frac{2}{\sqrt{6}} |0\rangle |-x\rangle = \frac{1}{\sqrt{6}} \sum_{j=1}^{4} (-1)^{j+1} |v_{j}\rangle ,
\end{equation}
satisfies these conditions.  Consequently, the POVM given by $\{ \Gamma (g)|X\rangle\langle X|\Gamma (g)^{-1} \, | \, g\in D_{3} \}$ will eliminate one of the six states $\{ \Gamma (g) |\psi_{e}\rangle \, | \, g\in D_{3} \}$.

\section{Eliminating more than one state}
So far, we have only explored measurements that eliminate one state, and now we would like to find one that eliminates sets of larger size.    Suppose we can find a vector $|X\rangle$, satisfying $\|X_{p}\|^{2}=d_{p}/|G|$, which is orthogonal to the states in $S_{e} = \{ |\psi_{e}\rangle , \Gamma (g_{1})|\psi_{e}\rangle , \ldots \Gamma (g_{n}) |\psi_{e}\rangle \}$.  We will then obtain a POVM whose outcomes correspond to eliminating the sets $S_{g} = \{ \Gamma (g)|\psi_{e}\rangle , \Gamma (gg_{1})|\psi_{e}\rangle , \ldots \Gamma (gg_{n}) |\psi_{e}\rangle \}$.  In general these sets may not be disjoint, and some may be identical.  If the group elements $\{ e, g_{1}, \ldots g_{n}\}$ form a subgroup, $H$, then the sets $S_{g}$ correspond to left cosets of $H$.  Any two cosets are either identical or disjoint, and there are $|G|/|H|$ of them, and that means that our POVM will be able to eliminate one of $|G|/|H|$ disjoint sets.  We will now look at two examples.

Let us consider three qubits, each of which is in one of the states $|\pm \theta\rangle$.  These eight states are generated by the group $\mathbb{Z}_{2} \times \mathbb{Z}_{2} \times \mathbb{Z}_{2}$ using the same representation of $\mathbb{Z}_{2}$ as before.  We will choose a vector $|X\rangle$ of the form
\begin{equation}
|X\rangle = \frac{1}{2\sqrt{2}} \sum_{j,k,l=0,1} e^{i\phi_{jkl}}|j\rangle\otimes |k\rangle \otimes |l\rangle ,
\end{equation}
and choose the phases so that $|X\rangle$ is orthogonal to both $|+\theta\rangle^{\otimes 3}$ and $|-\theta\rangle^{\otimes 3}$.  This will be true if 
\begin{eqnarray}
0 & = & e^{i\phi_{000}} + (e^{i\phi_{011}} + e^{i\phi_{101}} + e^{i\phi_{110}} )\tan^{2}\theta \nonumber \\
0 & = & (e^{i\phi_{001}} + e^{i\phi_{010}} + e^{i\phi_{100}} ) + e^{i\phi_{111}} \tan^{2}\theta .
\end{eqnarray}
If we choose $\phi_{011} = \alpha$, $\phi_{101}=0$, $\phi_{110}=-\alpha$, $\phi_{001}=\beta$, $\phi_{010}=0$, and $\phi_{100} = - \beta$, and both $\phi_{000}$ and $\phi_{111}$ equal to $\pi$, then these conditions, become
\begin{eqnarray}
1 & = & (1+2\cos\alpha ) \tan^{2}\theta \nonumber \\
\tan^{2}\theta & = & 1+2\cos\beta .
\end{eqnarray}
The first condition can be satisfied if $\tan^{2}\theta \geq 1/3$ and the second if $\tan^{2}\theta \leq 3$.  For $3 \geq \tan^{2}\theta \geq 1/3$ they can both be satisfied, and this determines a vector $|X\rangle$ that is orthogonal to both $|+\theta\rangle^{\otimes 3}$ and $|-\theta\rangle^{\otimes 3}$.  Note that $|-\theta\rangle^{\otimes 3} = R^{\otimes 3} |+\theta\rangle^{\otimes 3}$ and  that $\{ I^{\otimes 3}, R^{\otimes 3} \}$ is a subgroup.  That means the POVM will eliminate one of four sets, which in this case are pairs of states.  The elements of the pairs will differ in all three slots, that is, if, for example, the first element is $|+\theta , -\theta, -\theta\rangle$, the second element will be $|-\theta, +\theta , +\theta\rangle$.  So far, in our construction, we have $4$ pairs but $8$ POVM elements. The reason for this is that each pair is eliminated by two POVM elements.  For example, the pair $\{ |+\theta\rangle^{\otimes 3} , |-\theta\rangle^{\otimes 3} \}$ is eliminated by both $|X\rangle\langle X|$ and $(R\otimes R\otimes R)|X\rangle\langle X|(R\otimes R\otimes R)$, because $R\otimes R\otimes R$ maps the pair $\{ |+\theta\rangle^{\otimes 3} , |-\theta\rangle^{\otimes 3} \}$ into itself.  We can then combine these two rank one POVM elements into a rank two POVM element that eliminates the pair.  Doing the same thing with the remaining pairs, we finally have a $4$ element POVM, consisting of rank two operators, each of whose elements corresponds to eliminating a pair.  Note that there are $28$ possible pairs of states, so this POVM only eliminates a subset of the possible pairs. Also, due to the assumptions we made on $|X\rangle$, it is not clear whether the constructed measurement is optimal, or if it is possible to eliminate pairs also in other ranges of $\theta$.

This construction can be easily extended to $N$ states for each qubit, for $N$ even ($N$ needs to be even to guarantee that $|-\theta\rangle^{\otimes 3}$ is one of the states generated), by replacing $\mathbb{Z}_{2}$ by $\mathbb{Z}_{N}$, using the same representation for $\mathbb{Z}_{N}$ we used previously, and using the same vector $|X\rangle$, but with the factor $1/2^{3/2}$ replaced by $1/N^{3/2}$.  The result is an $N^{3}/2$ element POVM each of whose elements corresponds to eliminating a pair of states.

Moving on to four qubits, we can find a measurement that eliminates sets of four states.  The derivation is similar to those previously, so we will just state the results.  The vector $|X\rangle$ is chosen to be
\begin{equation}
|X\rangle = \sum_{j,k,l,m=0}^{1} z_{jklm} |jklm\rangle ,
\end{equation}
where $z_{1000}$, $z_{0010}$, $z_{0111}$, $z_{1101}$, $z_{0101}$, $z_{0110}$, $z_{1111}$ are equal to one, $z_{0011}=e^{i\alpha}$, $z_{1100} = e^{-i\alpha}$, and the remaining coefficients are equal to $-1$.  The angle $\alpha$ is given by 
\begin{equation}
\cos\alpha = \frac{1 - \tan^{4}\theta}{2\tan^{2}\theta} ,
\end{equation}
and this can be satisfied if $1 \geq \tan^{2}\theta \geq \sqrt{2}-1$.  This state is orthogonal to $|\theta\rangle^{\otimes 4}$, $|-\theta\rangle^{\otimes 4}$, $|\theta\rangle^{\otimes 2}|-\theta\rangle^{\otimes 2}$, and $|-\theta\rangle^{\otimes 2} |\theta\rangle^{\otimes 2}$.  The operators $\{ I^{\otimes 4}, R^{\otimes 4}, I^{\otimes 2}R^{\otimes 2}, R^{\otimes 2}I^{\otimes 2} \}$, which generate these states from $|\theta\rangle^{\otimes 4}$ form a subgroup.  Therefore, the sets that are eliminated are the four-element cosets of this subgroup, and there are four of them (the group is $\mathbb{Z}_{2}^{\times 4}$).  Each POVM element is four-dimensional and corresponds to eliminating one of the cosets.

Thus we see that we can use group theoretic methods to construct exclusion measurements that eliminate more than one state.  We next want to find a constraint on the kinds of exclusion measurements that are possible.

\section{Entropic bound}

Suppose Alice sends to Bob one of $N$ possible states, $|\psi_{z}\rangle$, $z\in \{ 1,2,\ldots N\}$, with all states being equally probable.  This set of states is divided into $M$ non-intersecting sets of size $K$, where $MK=N$.  Let $X\in \{ 1,2,\ldots M \}$ be the random variable corresponding to the set to which the state Alice sent belongs.  We denote the set of states corresponding to $X=x$ by $S_{x}$.  Bob performs a measurement on the state with $M$ possible outcomes, and $Y\in \{ 1,2,\ldots M\}$ is the random variable corresponding to Bob's outcome.  Each result of the measurement that Bob performs corresponds to eliminating one of the sets of size $K$ of the set of states, in particular, one of the $M-1$ sets to which the state that Alice sent does not belong.  This can be viewed as Alice sending one of the states
\begin{equation}
\rho_{x} = \frac{1}{K}  \sum_{z\in S_{x}} |\psi_{z}\rangle\langle \psi_{z}| ,
\end{equation}
where each $\rho_{x}$ is sent with a probability of $1/M$, and Bob's measurement yielding a result $y$, such that $\rho_{y}$ is not the state that Alice sent.  This scenario describes the two measurements in the previous section.

The mutual information between $X$ and $Y$ is (logarithms are base 2)
\begin{equation}
I(X:Y) = \sum_{x=1}^{M} \sum_{y=1}^{M} p(x,y) \log \left[ \frac{p(x,y)}{p_{X}(x) p_{Y}(y)} \right] ,
\end{equation}
where $p(x,y)$ is the joint distribution between $X$ and $Y$,  and $p_{X}(x)$ and $p_{Y}(y)$ are its marginals.  The measurement of this type that will provide the least information about which set the state that was sent belonged to (for a further discussion of this point see Appendix C) will be the one for which each each measurement result that can occur is equally probable, that is
\begin{equation}
\label{condmin}
p(y|x) = \left\{ \begin{array} {cc} 0 & y=x \\ 1/(M-1) & y \neq x\end{array} \right. ,
\end{equation}
where $p(y|x)$ is the  conditional probability of $y$ given $x$.  We already have that $p_{X}(x) = 1/M$, and from this and the above equation we have that $p_{Y}(y) = 1/M$.  The mutual information is then
\begin{equation}
\label{infbnd}
I(X:Y) = \log \left( \frac{M}{M-1} \right) .
\end{equation}

Any measurement that eliminates the same sets will have a greater mutual information that that given in Eq.\  (\ref{infbnd}).  Further, any measurement that eliminates the same sets will satisfy the Holevo  bound \cite{nielsen}, which implies that $I(X:Y) \leq S(\rho ) - \sum_{x=1}^{M} p_{X}(x) S(\rho_{x})$.  Here, if the state $\rho_{x}$ is sent with probability $p_{X}(x)$, then $\rho = \sum_{x=1}^{M} p_{X}(x) \rho_{x}$ (in our case we have that $p_{X}(x) = 1/M$).  .  Consequently, we have that 
\begin{equation}
\label{condition}
\log \left( \frac{M}{M-1} \right) \leq S(\rho ) - \frac{1}{M} \sum_{x=1}^{M} S(\rho_{x}) .
\end{equation}
This places a constraint on the sets of states for which it is possible to create an elimination measurement that will eliminate one of $M$ non-overlapping sets.

In the case that the states are generated by a group,  $\rho$ becomes quite simple, again with the caveat that each irreducible representation appears at most once, 
\begin{equation}
\rho = \frac{1}{|G|} \sum_{g\in G} \Gamma (g) |\psi_{e}\rangle\langle \psi_{e}| \Gamma (g)^{-1} = \sum_{p} \frac{1}{d_{p}} \| \psi_{ep}\|^{2} P_{p} ,
\end{equation}
where $|\psi_{ep}\rangle = P_{p}|\psi_{e}\rangle$.  We then have that 
\begin{equation}
S(\rho ) = - \sum_{p} \| \psi_{ep}\|^{2} \log (\| \psi_{ep}\|^{2}/d_{p}) .
\end{equation}
In the case of our $4$-qubit example., we have that
\begin{equation}
S(\rho ) = - \sum_{n=0}^{4} \left(\begin{array}{c} 4 \\ n \end{array} \right) s^{n} (1-s)^{4-n} \log [s^{n} (1-s)^{4-n}]  ,
\end{equation}
where $s=\sin^{2}\theta$. 

In our $4$-qubit example, the sets $S_{x}$ correspond to cosets of a subgroup, so the density matrices $\rho_{x}$ are related to each other by unitary operators, i.e.\ $\rho_{x^{\prime}} = \Gamma (g)\rho_{x}\Gamma (g)^{-1}$ for some $g\in G$.  This implies that they all have the same entropy.  Therefore we just have to find the entropy of the density matrix corresponding to the subgroup itself
\begin{eqnarray}
\rho_{e} & = & \frac{1}{4} [ ( |\theta\rangle\langle \theta |)^{\otimes 4} +( |\theta\rangle\langle \theta |)^{\otimes 2} (|-\theta\rangle\langle -\theta |)^{\otimes 2} \nonumber \\
 & & + ( |-\theta\rangle\langle -\theta |)^{\otimes 2}( |\theta\rangle\langle \theta |)^{\otimes 2}
+  ( |-\theta\rangle\langle -\theta |)^{\otimes 4} ].
\end{eqnarray}
This density matrix can be diagonalized, and its entropy is
\begin{eqnarray} 
S(\rho_{e}) & = & -\frac{1}{2} (1-v^{4}) \log (1-v^{4}) -\frac{1}{4} (1+v^{2})^{2} \log (1+v^{2})^{2} 
\nonumber \\ 
 & &  -\frac{1}{4} (1-v^{2})^{2} \log (1-v^{2})^{2} +2 ,
\end{eqnarray}
where $v=1-2s$.  With $M=4$ , the condition in Eq.\ (\ref{condition}) becomes
\begin{equation}
\log \left( \frac{4}{3} \right) \leq S(\rho ) - S(\rho_{e}) .
\end{equation}

\begin{figure} [h]
\includegraphics[scale=.15]{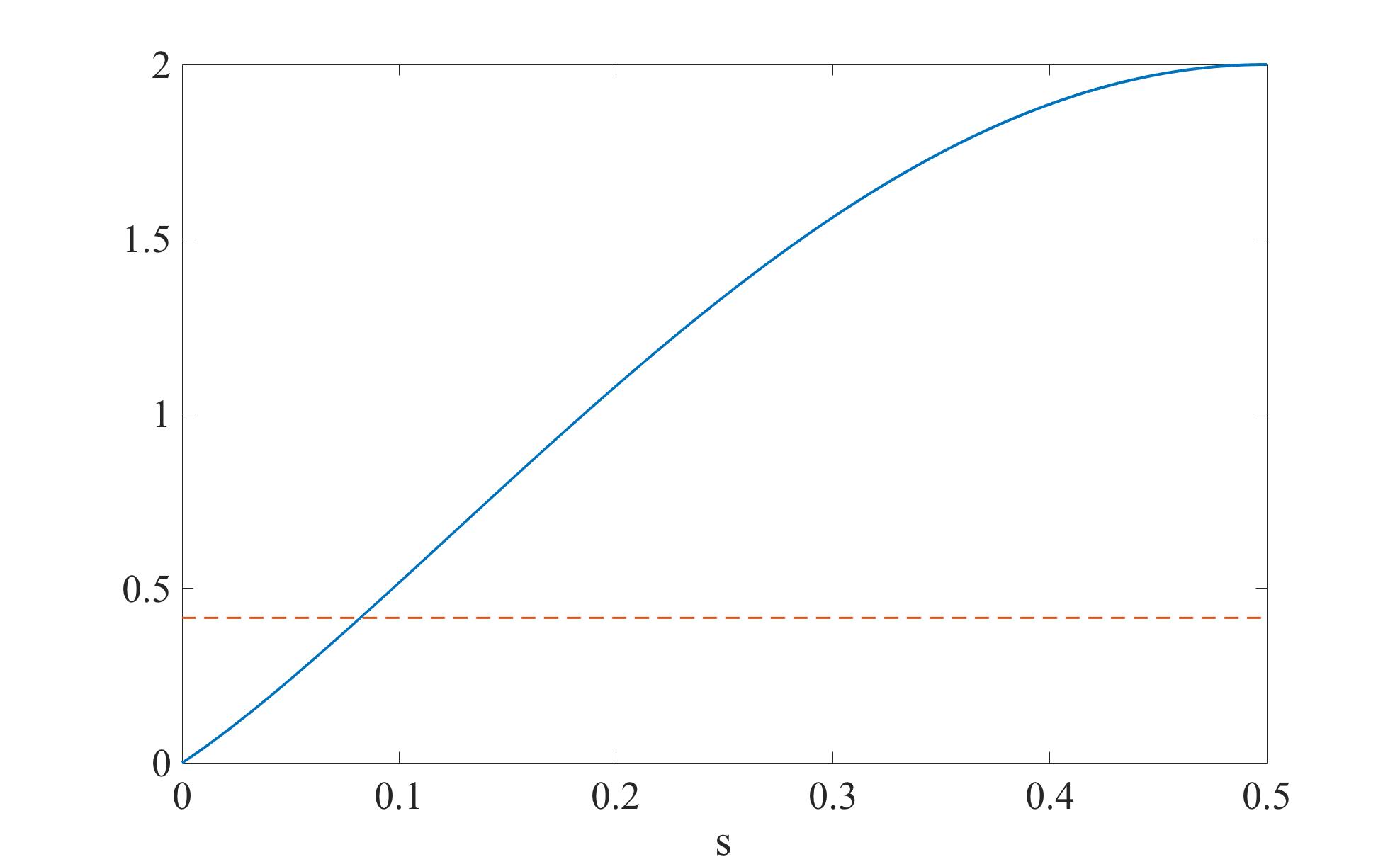}
\caption{Plot of the quantities appearing in Eq.\ (34) versus $s=\sin^{2}\theta$.  Green is the entropy difference and the red line is $\log (4/3)$.} 
\label{fig1}
\end{figure}

We plot the quantities in this inequality versus $s$ in Figure 1.  We see that for there to be a state elimination measurement that eliminates one of four sets, we must have $s>0.08$.  This implies that $\theta$ must be greater than $16$ degrees.

\section{Conclusion}
We have shown how group theory can be used to find measurements that eliminate states of qubit sequences.  We first looked at cases where one state is eliminated, a situation also known as anti-discrimination.  This was extended to cases in which more than one state is eliminated.  Finally, we developed a constraint on the construction of elimination measurements.

As was noted in the Introduction, elimination measurements have proven useful in a number of areas of quantum information.  By making it easier to find such measurements, we believe that the techniques presented here will increase the areas of applicability of these measurements.

\section*{Appendix A}

So far, we have only considered the situation in which the measurement always 
eliminates one of the states unambiguously.  This may not be possible in general.  We can extend the set of states for which elimination measurements are possible by allowing the measurement to sometimes fail, and telling us when it does.  As an example, let's go back to the case of two qubits, each in one of the states $|\pm \theta\rangle$, and see what we can do when $\tan\theta < \sqrt{2}-1$.  This will require a failure operator, that is a POVM element that will give us the probability of the measurement failing.  This case was studied in \cite{Crickmore}, but here we would like to consider from a slightly different point of view and include it for completelness.

We now set
\begin{equation}
|X\rangle = \sum_{j,k=0}^{1} c_{jk}|j\rangle\otimes |k\rangle ,
\end{equation}
and we still want the condition $\langle +\theta, +\theta |X\rangle = 0$, which is
\begin{equation}
\label{orthog}
c_{00}\cos^{2}\theta +(c_{01}+c_{10})\sin\theta \cos\theta + c_{11}\sin^{2}\theta = 0 .
\end{equation}
We will no longer have the condition that $|c_{jk}|$ is independent of $j$ and $k$, because we cannot satisfy the above equation if it holds.  The POVM operators that eliminate a state are still given by $\Pi_{g} = \Gamma (g) |X\rangle\langle X| \Gamma(g)^{-1}$, and the failure operator $\Pi_{f}$ is given by
\begin{equation}
\Pi_{f}= I - \sum_{g} \Pi_{g} =I - 4\sum_{j,k=0}^{1} |c_{jk}|^{2} |j\rangle\langle j| \otimes |k\rangle \langle k|.
\end{equation}
For $\Pi_{f}$ to be a positive operator, we see from the above equation that we must have $|c_{jk}| \leq 1/2$.  Assuming the states are equally likely, the failure probability is
\begin{eqnarray}
P_{f} & = & \frac{1}{4} \sum_{j,k=\pm \theta} \langle j,k|\Pi_{f}|j,k\rangle \nonumber \\
& = & 1- 4( |c_{00}|^{2} \cos^{4}\theta + ( |c_{01}|^{2} + |c_{10}|^{2}) \sin^{2}\theta \cos^{2}\theta \nonumber \\
& & + |c_{11}|^{2} \sin^{4}\theta ) .
\end{eqnarray}
We want to minimize $P_{f}$, which means we want to maximize the expression in parentheses in the above equation.  The coefficient multiplying $|c_{00}|^{2}$ is the largest, so we would like to make $|c_{00}|$ as large as possible consistent with the condition in Eq.\ (\ref{orthog}).  Now if we choose $c_{00}$ real and positive, looking at Eq.\ (\ref{orthog}), we see it will be maximized if we choose $c_{01}=c_{10}=c_{11}= -1/2$.  This then gives us
\begin{equation}
c_{00}=\tan\theta + \frac{1}{2} \tan^{2}\theta ,
\end{equation}
and the condition $\tan\theta < \sqrt{2}-1$ guarantees that $|c_{00}| < 1/2$.  This, then, specifies the POVM elements, and the failure probability is given by
\begin{equation}
P_{f}=1- 2 \sin^{2}\theta [ 1+2\cos\theta\, (\cos\theta + \sin\theta )] .
\end{equation}
Note that this expression holds only for $\theta \leq \pi /8$, and $P_{f}=0$ for $\theta \geq \pi /8$. Again, we have not proven that this is the optimal success probability, but it turns out that it is \cite{Crickmore}.  Finally, if one goes from $\mathbb{Z}_{2} \times \mathbb{Z}_{2}$ to $\mathbb{Z}_{N} \times \mathbb{Z}_{N}$ using the same representation as in the previously, the expressions for the failure operator and the failure probability remain the same.

\section*{Appendix B}
So far, we have only considered two-,  three-, and four-qubit states, but in \cite{PBR} measurements that exclude a single $n$-qubit state were found.  It is useful to study this case from the group theory point of view.  Let us consider the set of $n$-qubit states where each qubit is in either the state $|+\theta\rangle$ or the state $|-\theta\rangle$.  We will denote a member of this set as $|\Psi_{x}\rangle$, where $x$ is an $n$-digit binary number, and a $0$ in the $j^{\rm th}$ place corresponds to the $j^{\rm th}$ qubit being in the state $|+\theta\rangle$ and a $1$ corresponds to the $j^{\rm th}$ qubit being in the state $|-\theta\rangle$.  We want to find a measurement that will eliminate one of the states $|\Psi_{x}\rangle$.

The relevant group here is $\mathbb{Z}_{2}^{\times n}$ and the representation for $\mathbb{Z}_{2}$ is the same one we used before.  The invariant subspaces for the irreducible representations are just the basis vectors $|x\rangle = \prod_{j=0}^{n-1} |x_{j}\rangle$, with each $x_{j}$ being either $0$ or $1$.  That means that the vector $|X\rangle$ is given by
\begin{equation}
|X\rangle = \frac{1}{2^{n/2}} \sum_{x=0}^{2^{n}-1} e^{i\phi_{x}} |x\rangle ,
\end{equation}
The condition $\langle \Psi_{0} | X\rangle = 0$ gives us
\begin{equation}
\sum_{x=0}^{2^{n}-1} e^{i\phi_{x}} \langle \Psi_{0}|x\rangle = 0.
\end{equation}
Let us now make the Ansatz $\phi_{0} = \alpha$ and $\phi_{x} = |x|\beta$ for $x\neq 0$, where $|x|$ is the number of ones (Hamming weight) in the sequence $x$.  We then have
\begin{equation}
e^{i\alpha} \cos^{n}\theta + \sum_{k=1}^{n} \left[ \left(\begin{array}{c} n \\ k \end{array}\right) e^{ik\beta}\cos^{n-k}\theta \sin^{k}\theta \right] = 0,
\end{equation}
or, factoring out $\cos^{n}\theta$,
\begin{equation}
e^{i\alpha} + (1 + e^{i\beta} \tan\theta )^{n} -1 = 0.
\end{equation}
This is the condition derived in \cite{PBR}, and it was shown there that it is possible to satisfy it for $\arctan (2^{1/n} -1) \leq \theta \leq \pi /4$. 

As we did with two qubits, we can increase the number of states for each qubit to $N$, and the group to $\mathbb{Z}_{N}^{\times n}$, using the same two dimensional representation of $\mathbb{Z}_{N}$ as before.  The same vector $|X\rangle$, which we just found, with $1/2^{n/2}$ replaced by $1/N^{n/2}$, can be used to form the POVM.  Thereby we obtain a POVM that will eliminate one state from sequences of $n$ qubits, where each qubit can be in one of $N$ states.  

\section*{Appendix C}

Here we want to show that the conditional probability given in Eq.\ (\ref{condmin}) does lead to a minimum of the mutual information.  We assume that we have a measurement that eliminates one of $M$ non-overlapping sets, with each state being equally probable, but that 
\begin{equation}
p(y|x) = \left\{ \begin{array} {cc} 0 & y=x \\ q_{yx} & y \neq x \end{array} \right. ,
\end{equation}
where $0\leq q_{yx} \leq 1$ and they satisfy the $M$ constraints
\begin{equation}
\sum_{ \{ y| y \neq x\} } q_{yx} = 1 .
\end{equation}
We then find that
\begin{equation}
\label{pY}
p_{Y}(y) = \frac{1}{M} \sum_{x \neq y} q_{yx} ,
\end{equation}
and
\begin{eqnarray}
I(X:Y) & = & -\frac{1}{M} \sum_{x=1}^{M} \sum_{ y\neq x } q_{yx} \log q_{yx} \nonumber \\
 & & + \sum_{y=1}^{M} p_{Y}(y) \log p_{Y}(y) .
\end{eqnarray}
For each constraint we have a Lagrange multiplier, $\lambda_{x}$, and we then have the equations
\begin{equation}
\frac{\partial}{\partial q_{yx}} \left[ I(X:Y) - \sum_{x^{\prime} = 1}^{M} \sum_{ y^{\prime} \neq x^{\prime}}\lambda_{x^{\prime}} q_{y^{\prime}x^{\prime}} \right] = 0
 \end{equation}
 This gives 
 \begin{equation}
\log  q_{yx} = M\lambda_{x} + \log p_{Y}(y)  ,
\end{equation}
or $q_{yx}= e^{M\lambda_{x}} p_{Y}(y)$.  If we now insert this expression for $q_{yx}$ into Eq.\ (\ref{pY}), we get the consistence condition
\begin{equation}
\frac{1}{M} \sum_{x\neq y} e^{M\lambda_{x}} = 1 .
\end{equation}
For this to hold for each value of $y$, it must be the case that $e^{M\lambda_{x}}$ is a constant, independent of $x$, and, therefore, equal to $M/(M-1)$.  If we now insert this into the constraint equations, we find, for each $x$
\begin{equation}
\frac{M}{M-1} \sum_{y\neq x} p_{Y}(y) = 1 ,
\end{equation} 
For this to hold $p_{Y}(y)$ must be constant, and equal to $1/M$.  This then yields $q_{yx}= 1/(M-1)$ for $y \neq x$.

\acknowledgments
This work was supported by the UK Engineering and Physical Sciences Research Council (EPSRC) under EP/M013472/1 and  EP/L015110/1.

\end{document}